\documentclass[12pt]{article}\textheight 24.5cm \textwidth 17cm\voffset=-1.in\hoffset= - 1.in

\makeatletter
\@addtoreset{equation}{section}

\makeatother

\usepackage[utf8]{inputenc}
\usepackage[T2A]{fontenc}
\usepackage{amsmath}
\usepackage{slashed}
\usepackage{graphicx}
\usepackage{amsfonts}
\usepackage{amssymb}
\usepackage{psfrag}
\usepackage{hhline}
\usepackage{cite}
\usepackage{upgreek}
\usepackage{epsfig,amsfonts}
\usepackage{feynmp}

\author{ Lev Spodyneiko\footnote{E-mail:  lionspo@itp.ac.ru}
\vspace*{10pt}\\[\medskipamount]
\parbox[t]{0.88\textwidth}{\normalsize\it\raggedright
L.D.Landau Institute for Theoretical Physics,
142432 Chernogolovka, Russia}
}
\date{}
\title{AGT сorrespondence, Ding--Iohara algebra at  roots of unity and Lepowsky--Wilson construction}

\begin{document}

\maketitle

\begin{abstract}

 It was recently conjectured that the AGT correspondence between the $U(r)$--instanton counting on $\mathbb R^4/\mathbb Z_p$ and the two-dimensional field theories with the conformal symmetry algebra $\mathcal A(r,p)$ can be considered as a root of unity limit of its K-theoretic analogue. From this point of view, the algebra  $\mathcal A(r,p)$ and a special basis in its representation are limits of the Ding--Iohara algebra and the Macdonald polynomials respectively. In this paper we confirm this conjecture for the special case $r=1$. We uncover the implicit $\mathcal A(1,p)$ symmetry in this limit. We also found that the vertex operators in the special basis have factorized AFLT form.
\end{abstract}
\section{Introduction}

In the paper \cite{Alday:2009aq} Alday, Gaiotta and Tachikawa proposed a relation between the two-dimensional conformal field theories and $\mathcal N = 2$ four-dimensional SUSY gauge theories. This conjecture is usually called the AGT correspondence. It provides a combinatorial formula for the expansion of conformal blocks by using an exact form for the Nekrasov instanton partition function. In \cite{Alba:2010qc} AGT correspondence was understood in terms of matrix elements of vertex operators. It was discovered that there is a special orthogonal basis in the representation of the algebra $\mathcal A(1,2) = \mathcal H \oplus \mathsf{Vir}$. This basis is uniquely defined by a property that matrix elements of primary fields have a simple factorized form. Moreover, this basis diagonalizes a system of integral of motions related to the Benjamin--Ono hierarchy.

 Generalizations of the AGT-relation was suggested in \cite{Belavin:2011pp,Nishioka:2011jk,Belavin:2011sw}. It was conjectured that conformal blocks of the coset construction
 \begin{equation} \label{A(r,p) algebra}
 \mathcal A(r,p) = \mathcal H \otimes \widehat {\mathfrak{sl}}(p)_r \otimes \frac{\widehat {\mathfrak{sl}}(p)_p \otimes \widehat {\mathfrak{sl}}(p)_{n-p}}{\widehat {\mathfrak{sl}}(p)_n}
 \end{equation}
correspond to the instanton partition function of the $U(r)$ gauge theory on $\mathbb{R}^4/\mathbb Z_p$. Note, that in the special case  $r=1$ this coset simplifies to
\begin{equation} \label{A(1,p) algebra}
\mathcal A (1,p) = \mathcal H \oplus \widehat {\mathfrak {sl}}(p)_1.
\end{equation}

On the other hand, Awata et al.\ proposed in \cite{Awata:2009ur,Awata:2011dc} a $q$-analagues version of the AGT-relation for the $U(r)$ instantons on $\mathbb R^4$ case. They suggested that the level $m$ representation of the Ding--Iohara algebra \cite{ShiffVass,Feigin:2013fga,Awata:2011dc} have a special basis described by the Macdonald symmetric functions in which the vertex operators have factorized form. Furthermore, there is an infinite set of integrals of motion corresponding to the Macdonald difference operators. The Ding--Iohara algebra, its representations and the Macdonald polynomials depend on two arbitrary parameters $q,t$ and in the limit $q,t \rightarrow 1$ the results of non-deformed AGT-relation are recovered.

As $q,t \rightarrow 1$ the Macdonald polynomials tend to the Jack polynomials which are in a one-to-one correspondence with the special basis of the work \cite{Alba:2010qc}. On the other hand, Uglov \cite{Uglov:1997ia,Takemura:1996qv} had shown that the Macdonald polynomials in the limit $q,t \rightarrow e^{2\pi i /p }$ correspond to the solutions of the spin generalization of the Calogero-Sutherland model. The model was also found to possess the $\mathcal H \oplus \hat{\mathfrak{sl}}(p)_1$ symmetry. The fact that this symmetry coincides with the conformal algebra $\mathcal A(1,p)$ lead the authors of \cite{Belavin:2012eg} to the following conjecture. It was argued that in the $q,t \rightarrow e^{2\pi i /p }$ limit the level $r$  representation of the Ding--Iohara algebra tends to the conformal algebra $\mathcal A(r,p)$ and the special basis in the representation of  the $\mathcal A(r,p)$ algebra is the limit of the corresponding basis in  the Ding--Iohara algebra. The conjecture was only proved \cite{Belavin:2012eg} for the particular case of the $\mathcal A(1,2)$ algebra and some non-trivial checks for the case of $\mathcal A(2,2)$ were performed. See also \cite{Itoyama:2013mca,Itoyama:2014pca}.

In this paper we extend the results of \cite{Belavin:2012eg}. Namely, we study the AGT relation in the case of the algebra $\mathcal A(1,p)$. We consider limit $q,t \rightarrow e^{2\pi i /p }$ of the level 1 representation of the Ding--Iohara algebra. In this limit the $\mathcal A(1,p)$ symmetry is not manifested and we explicitly construct it. It turns out that in this limit the algebra $\mathcal A(1,p)$ naturaly arises in its Lepowsky--Wilson realisation \cite{GLW}. We also present the formulas for the matrix elements of the vertex operators and show that the character of $\mathcal A(1,p)$ is equal to the generating function of the colored Young diagrams.

 The plan of the paper is the following. In section 2 we review the Lepowsky--Wilson construction. In section 3 we consider the root of unity limit of the Ding--Iohara algebra and uncover its implicit $\mathcal A(1,p)$ symmetry.  In section 4 we construct the basis and show that the vertex operators have factorised form in it. In section 5 we study a relation between  the generating function of colored Young diagrams and the character of the representation of the algebra $\mathcal A(1,p)$.  Appendix A contains the basic definitions of partitions and of the Macdonald polynomials.

\section{Lepowsky--Wilson construciton}
\subsection{Lepowsky--Wilson construction for $\widehat{\mathfrak {sl}}(2)_1$}
In this section we review the Lypowsky--Wilson construction for the $\widehat{\mathfrak{sl}}(2)_1$ algebra as it was originally introduced in \cite{Lepowsky:1978jk}. We choose the generator $e_n,f_n,h_n$ in $\widehat{\mathfrak{sl}}(2)_1$ with the commutation relations
\begin{align}
\begin{split}
[h_n, e_m] = 2e_{n+m}, \quad [h_n,f_m] = -2f_{n+m},\quad [h_n,h_m] = 2n\delta_{n+m,0},\\
[e_n,f_m] = h_{n+m} +n\delta_{n+m},\quad [e_n,e_m] = [f_n,f_m] = 0,
\end{split}
\end{align}
It is known that the level one integrable representation of this algebra can be realised in terms of its Heisenberg subalgebra generated by $h_n$. This is called the Frenkel--Kac construction. On the other hand, there is another Heisenberg subalgebra generated by the operators $a_{2n+1}\equiv e_n+f_{n+1}$ and the level one integrable representation of $\widehat{\mathfrak{sl}}(2)_1$ can be realised as its Fock space. Let us choose a different basis in $\widehat{\mathfrak{sl}}(2)$
\begin{equation}\label{LW p=2}
a_{2n+1} = e_n+f_{n+1}, \quad b_{2n+1} =e_n- f_{n+1}, \quad b_{2n} =- h_n + \frac 1 2 \delta_{n} \hat k.
\end{equation}
These generators obey the relations
\begin{align}\label{LW comm p=2}
\begin{split}
[a_{2n+1},a_{2m+1}] = (2n+1) \delta_{n+m+1}, \quad [a_{2n+1},b_m] = 2 b_{m+2n+1}, \quad [b_{2n},b_{2m}] = 2n \delta_{n+m}\\
[b_{2n+1},b_{2m+1}] = - (2n+1) \delta_{n+m+1}, \quad [b_{2n+1},b_{2m}] = 2 a_{2m+2n+1}.
\end{split}
\end{align}
Lepowsky and Wilson found that in the level one representation action of the operators $b_n$ is the same as of exponents of the generators $a_{2n+1}$:
\begin{equation}\label{p=2 LW}
b(z) = \sum b_n z^{-n} = \frac {(-1)^{\sigma}} 2 \exp \left( 2 \sum_{n\ne 0} \frac {a_{2n+1}}{-2n-1} z^{-2n-1}\right),
\end{equation}
where the variable $\sigma=0,1$ labels different level one integrable representations $\pi_{\sigma,1}$ of $\widehat{\mathfrak{sl}}(2)$. We will prove that these exponents indeed obey (\ref{LW comm p=2}) later in this paper.
\subsection{Lepowsky--Wilson construction for the $\widehat{\mathfrak {sl}}(3)_1$}
 Now, we consider the case of the $\widehat{\mathfrak {sl}}(3)_1$ algebra. We will use the matrix basis in $\widehat{\mathfrak {sl}}(3)$. The algebra is generated by the operators $E^{a,b}_n$, where $a,b= 1,2,3$ and $n\in \mathbb Z$. They satisfy the relations
 \begin{align}
[E^{a,b}_n, E^{c,d}_m] = \delta_{b,c} E^{a,d}_{n+m} - \delta_{a,d} E^{c,a}_{n+m} +  \delta_{n+m,0} \delta_{b,c} \delta_{a,d},
\end{align}
 These operators generate the $\widehat{\mathfrak {gl}}(3)_1$ algebra. The $\widehat{\mathfrak {sl}}(3)_1$ algebra is generated by their traceless combinations.

 To make connection with the previous section, note that a similar basis in $\widehat{\mathfrak {sl}}(2)_1$ consists of the elements $E^{a,b}_n$ with $a,b=1,2$, $n \in \mathbb Z$ so that $e_n = E^{1,2}_n$, $ f_n= {E^{2,1}_n}$, $h_n = E^{1,1}_n-E^{2,2}_n$.

 Again, there is a special Heisenberg subalgebra with the generators, whuch will be denoted as $a_{3n+1}, a_{3n+2}$, and the level one representation of $\widehat{\mathfrak {sl}}(3)_1$ can be realised as the Fock module of this subalgebra. Define another basis in $\widehat{\mathfrak {sl}}(3)_1$

\begin{equation} \begin{split}
 \begin{aligned}
a_{3n+1}&= E^{23}_n+E^{31}_{n+1}+E^{12}_n,&
a_{3n+2}&= E^{13}_n+E^{21}_{n+1}+E^{32}_{n+1},\\
b^{(1)}_{3n+1}&= E^{23}_n+\overline\omega E^{31}_{n+1}+{\overline\omega}^2 E^{12}_n,&
b^{(1)}_{3n+2}&= E^{13}_n+\omega E^{21}_{n+1}+\omega^2 E^{32}_{n+1},\\
b^{(2)}_{3n+1}&= E^{23}_n+\omega E^{31}_{n+1}+{\omega}^2 E^{12}_n,&
b^{(2)}_{3n+2}&= E^{13}_n+\overline\omega E^{21}_{n+1}+\overline\omega^2 E^{32}_{n+1},\\
b^{(1)}_{3n}&= E^{33}_n+ \overline \omega E^{11}_n +\overline \omega^2 E^{22}_n + \frac{1}{1-\omega} \delta_{n,0},&
b^{(2)}_{3n}&= E^{33}_n+ \omega E^{11}_n + \omega^2 E^{22}_n + \frac{1}{1-\overline\omega} \delta_{n,0},
\end{aligned}
\end{split}
\end{equation}
where the variable $\omega = e^{2\pi i/3}$,  $\overline\omega = e^{-2\pi i/3}$. The action of the operators $b^{(1)}_n$ and $b^{(2)}_n$ can be realised in terms of the exponents of $a_{3n+1},a_{3n+2}$. Namely,
\begin{align}
&b^{(1)}(z) = \sum_n b^{(1)}_n z^{-n} =\frac{\omega^{2\sigma}}{1-\omega} \exp\Big((1-\omega)\!\!\!\!\sum_{n\equiv 1 \!\!\!\!\mod 3} \frac{a_{-n}}{n}z^n+ (1-\overline\omega)\!\!\!\!\sum_{n \equiv 2 \!\!\!\!\mod 3} \frac{a_{-n}}{n}z^n\Big),\\
&b^{(2)}(z)= \sum_n b^{(2)}_n z^{-n} =\frac{\omega^\sigma}{1-\overline\omega} \exp\Big((1-\overline \omega)\!\!\!\!\sum_{n\equiv 1 \!\!\!\!\mod 3} \frac{a_{-n}}{n}z^n+ (1-\omega)\!\!\!\!\sum_{n \equiv 2 \!\!\!\!\mod 3} \frac{a_{-n}}{n}z^n\Big),
\end{align}
where $\sigma=1,2,3$ label different level one integrable representations $\pi_{\sigma,1}$ of $\widehat{\mathfrak{sl}}(3)$.

\subsection{Generalized Lepowsky--Wilson construction}
Finally we present the Lepowksy--Wilson construction \cite{GLW} for the algebra $\widehat{\mathfrak {sl}}(p)_1$ with general~$p$.

We will choose the matrix basis in $\hat {\mathfrak{sl}}(p)_1$. The algebra is generated by operators $E_n^{a,b}$ with $a,b=1,\dots, p$ and $n\in \mathbb Z$. They obey
\begin{align}
[E^{a,b}_n, E^{c,d}_m] = \delta_{b,c} E^{a,d}_{n+m} - \delta_{a,d} E^{c,a}_{n+m} +  \delta_{n+m,0} \delta_{b,c} \delta_{a,d},
\end{align}
 These elements generate the algebra $\hat {\mathfrak{gl}}(p)_1$, their  traceless combinations generate the algebra $\hat {\mathfrak{sl}}(p)_1$.
 It will be useful in what follows to regard the indexes $a,b$ in $E^{a,b}_n$ as numbers modulo $p$ (e.g. $E^{p+a,b}_n \equiv E^{a,b}_n$).

Define  another basis
\begin{align}\label{lw relation}
b^{(s)}_{p n + k} = \sum_{r=0}^{p-1} E^{-k+r,r}_{n+\theta(-k+r,r)} \omega_p^{-rs} + \frac{1}{1-\omega^{s}_p}\delta_{n,0}\delta_{k,0}.
\end{align}
Here $\theta(a,b)$ is equal $1$ if $E^{a,b}$ is strictly upperdiagional, and $0$ otherwise, $\omega_p = e^{2\pi i/p}$ , $k=0,\dots,p-1$,  in the case $s=0$ the last term should be omitted. Note, that in the previous sections we used notation $a_n$ for $b^{(0)}_n$.

By  straightforward calculation
\begin{align*}
\begin{split}
&[b^{(s)}_{p n + k},b^{(t)}_{p n + l}]= \sum_{r,r'} [ E^{-k+r,r}_{n+\theta(-k+r,r)}, E^{-l+r',r'}_{m+\theta(-l+r',r')}] \omega_p^{-rs -r't}\\
&= \sum_{r,r'} \left(\delta^{(p)}_{r,-l+r'} E^{-k+r,r'}_{n+m+\theta(-k+r,r')}-\delta_{r',-k+r}E^{-l+r',r}_{n+m+\theta(-l+r',r)}\right.\\&+\left.(n+\theta(-k+r,r)) \delta^{(p)}_{-k+r,r'}\delta_{r,-l+r'}^{(p)}\delta_{n+m+1}\right)\omega_p^{-rs-r't}\\&=n  \omega^{-sn}_p\delta_{n+m,0} \delta^{(p)}_{s+t,0} + (\omega^{sm}_p - \omega^{tn}_p)b^{(s+t)}_{n+m},
\end{split}
\end{align*}
we see that the generators $b^{(s)}_n$ satisfy the commutation relations
\begin{align}\label{b comm LW}
[b^{(s)}_n,b^{(t)}_m] &= n  \omega^{-sn}_p\delta_{n+m,0} \delta_{s+t,p} + (\omega^{sm}_p - \omega^{tn}_p)b^{(s+t)}_{n+m}.
\end{align}
We will show in what follows that  the level one representation of $\hat {\mathfrak{sl}}(p)_1$ can be realized as Fock module of the Heisenberg subalgebra $a_n = b^{(0)}_n$ with $n\not \equiv 0\mod p$.  The action of the generators $b^{(s)}_n$ on this module is defined as

\begin{align}
&b^{(s)}(z) =\sum_n b^{(s)}_n z^{-n}=\frac{\omega_p^{\sigma(p-s)}}{1-\omega_p^s} :\exp\big( \sum_{n\in \mathbb Z \backslash \{0\}} \frac{a_n}{-n} z^{-n}(1- \omega_p^{-sn}) \big):, \qquad \text{for $s=1,2,\dots,p-1$}.
\end{align}

\section{Ding--Iohara algebra at roots of unity}
\subsection{Ding--Iohara algebra}

The Ding--Iohara algebra (other names are  {\it quantum toroidal} $\mathfrak{gl}_1$ or elliptic Hall algebra)\cite{ShiffVass,Feigin:2013fga,Awata:2011dc} is associative algebra generated by the currents $x^{\pm}(z) = \sum_{n\in \mathbb Z} x_n^{\pm} z^{-n}, \psi^{\pm} = \sum_{\pm n\in \mathbb Z_{\geq 0}} \psi^{\pm}_n z^{-n}$ and the central element $\gamma^{\pm1/2}$. The currents satisfy the relations
\begin{equation*}
\begin{aligned}
&\psi^\pm(z) \psi^\pm(w) = \psi^\pm(w) \psi^\pm(z),
\\ &\psi^+(z)\psi^-(w) =
 \dfrac{g(\gamma^{+1} w/z)}{g(\gamma^{-1}w/z)}\psi^-(w)\psi^+(z),
\\
&\psi^+(z)x^\pm(w) = g(\gamma^{\mp 1/2}w/z)^{\mp1} x^\pm(w)\psi^+(z),
\\
 &\psi^-(z)x^\pm(w) = g(\gamma^{\mp 1/2}z/w)^{\pm1} x^\pm(w)\psi^-(z),
\\
&[x^+(z),x^-(w)]
 =\dfrac{(1-q)(1-1/t)}{1-q/t}
 \big( \delta(\gamma^{-1}z/w)\psi^+(\gamma^{1/2} w)-
 \delta(\gamma z/w)\psi^-(\gamma^{-1/2}w) \big),
\\
&G^{\mp}(z/w)x^\pm(z)x^\pm(w)=G^{\pm}(z/w)x^\pm(w)x^\pm(z),
\end{aligned}
\end{equation*}
where
\begin{align*}
\delta(z)=\sum_{n\in \mathbb Z } z^n, \qquad g(z) \equiv \dfrac{G^+(z)}{G^-(z)},\quad
G^\pm(z) \equiv (1-q^{\pm 1}z)(1-t^{\mp 1}z)(1-q^{\mp 1}t^{\pm 1}z).
\end{align*}
This algebra depends on two complex parameters $q,t$, and we will denote it by $\mathcal E_1 (q,t)$.
Its representation is said to be of the level $m$, if the central element takes the value $\gamma^{\pm 1/2} = (t/q)^{\pm m/4}$.

\subsection{Level one representation of Ding--Iohara algebra.}

The level one representation of the Ding--Iohara algebra could be constructed as a module of a deformed Heisenberg algebra $\mathcal H(q,t)$ \cite{Awata:2011dc}. $\mathcal H(q,t)$ consist of the generators $\{a_n \mathbin{|}n\in\mathbb Z\backslash\{0\}\}$ which satisfy the commutation relations
\begin{align}
[a_n,a_m] = n \frac{1-q^{|m|}}{1-t^{|m|}} \delta_{m+n,0}.
\end{align}
The Fock space is defined by its vacuum state $|0\rangle$ satisfying the annihilation condition $a_n |0\rangle =0$ for $n > 0$.

The level 1 representation of the Ding--Iohara algebra can be realised as the Fock module by means of the formulas
\begin{align}\label{level 1 repr}
\begin{split}
& \gamma^{\pm1/2} \mapsto (t/q)^{\pm1/4} \\
&x^+(z)\mapsto
u\exp\Big( \sum_{n=1}^{\infty} \dfrac{1-t^{-n}}{n}a_{-n} z^{n} \Big)
\exp\Big(-\sum_{n=1}^{\infty} \dfrac{1-t^{n} }{n}a_n    z^{-n}\Big),\\
&x^-(z)\mapsto u^{-1}
\exp\Big(-\sum_{n=1}^{\infty} \dfrac{1-t^{-n}}{n}(t/q)^{n/2}a_{-n} z^{n}\Big)
\exp\Big( \sum_{n=1}^{\infty} \dfrac{1-t^{n}}{n} (t/q)^{n/2} a_n z^{-n}\Big),\\
&\psi^+(z)\mapsto
\exp\Big(
 -\sum_{n=1}^{\infty} \dfrac{1-t^{n}}{n} (1-t^n q^{-n})(t/q)^{-n/4} a_n z^{-n}
    \Big),
\\
&\psi^-(z)\mapsto
\exp\Big(
 \sum_{n=1}^{\infty} \dfrac{1-t^{-n}}{n} (1-t^n q^{-n})(t/q)^{-n/4} a_{-n}z^{n}
    \Big).
\end{split}
\end{align}
We will call this module $\mathcal F_u$.

\subsection{$\omega_p$ limit of level 1 representation.} \label{vert limit}
It was conjectured \cite{Belavin:2012eg} that in the limit $q,t \rightarrow e^{2\pi i / p}$ a level $r$ representation of the Ding--Iohara algebra tends to a representation of the conformal algebra $\mathcal A(r,p)$. It was proved for the case of $\mathcal A(1,2)$ and checked for $\mathcal A(2,2)$ in \cite{Belavin:2012eg}. In this paper we are interested in the case of the  $\mathcal A(1,p)$ algebra. Therefore, we consider the following limit of the level 1 representation of the Ding--Iohara algebra
\begin{equation}\label{limit wp}
t= \omega_p e^{\beta\hbar },\quad q = \omega_p e^{\hbar},\quad u =\omega_p^\sigma e^{\kappa \hbar},\quad \omega_p = e^{\frac{2\pi i}{p}},\quad \hbar\rightarrow 0
\end{equation}
in the representation (\ref{level 1 repr}). Here $\sigma= 0,\dots p-1$ labels different representations.

Note, that commutation relation of $a_n$ in this limit are
\begin{equation}
[a_n,a_m] =  \left\{ \begin{aligned} &n \delta_{m,n},&& n \ne 0 \mod p , \\  &\frac 1  \beta n \delta_{m,n},&&  n=0 \mod p.\end{aligned} \right.
\end{equation}
We will replace $a_{pn} \rightarrow \frac 1 {\sqrt \beta} a_{pn}$ to get the standard Heisenberg commutators
\begin{equation}
[a_n,a_m] = n \delta_{m,n}.
\end{equation}

Define
\begin{align}\label{b limit}
&\Phi(z) = -\sum_{n\in \mathbb Z \backslash \{0\}} \frac{a_n}{n} z^{-n},\\
&b^{(s)}(z) = \frac{1}{1-\omega_p^s} :\exp\big( \Phi(z)-\Phi(\omega_p^s z) \big):, \qquad \text{for $s=1,2,\dots,p-1$}
\end{align}
where $:\,:$ means the normal ordering. Notice that in the limit (\ref{limit wp}) relations (\ref{level 1 repr}) become
\begin{align*}
&x^{\pm} =  b^{(p-1)}(z) + O(\hbar).
\end{align*}

These fields have the following operator product expansions (OPEs)
\begin{align}
\Phi(u)\Phi(u)\sim \ln(u-z)
\end{align}
\begin{align}\label{bb ope}
b^{(s)}(u)b^{(t)}(z)\sim\frac{1}{(1-\omega^s)(1-\omega^t)} \frac{(u-z)(u-\omega^{t-s}z)}{(u-\omega^{-s} z)(u-\omega^t z)}:\exp\big( \Phi(u)-\Phi(\omega^s u)+ \Phi(z)-\Phi(\omega^t z)\big):.
\end{align}
From this OPE we can obtain the commutation relations for the modes $b^{(s)}(z)=\sum_{n\in \mathbb Z} b_n^{(s)} z^{-n}$. Let us derive, for example, the commutator $[b^{(s)}_n,b^{(t)}_m]$ with $s+t \ne p$. To get it we have to multiply (\ref{bb ope}) by $u^{n-1} z^{m-1}$ and integrate over $z$ along the circle contour $C_1$ around zero, then to integrate over $u$ along the larger circle $C_2$ and, at last, to subtract the same done in the opposite order. This reads
\begin{align}
\left(\oint_{C_2} du\,  u^{n-1}\oint_{C_1}dz \,z^{m-1} -\oint_{C_2} dz\, z^{m-1}\oint_{C_1}du\,  u^{n-1}  \right)  b^{(s)}(u)b^{(t)}(z).
\end{align}
In the first term, by deforming the $C_2$ countour through the poles  $u= \omega^{-s} z$ and $u=\omega^{t}z$, we obtain
\begin{align}
\begin{split}
\oint_{C_1}dz z^{n+m-1}\left(\frac{\omega^{-sn}}{1-\omega^{s+t}} \exp\big( \Phi(\omega^{-s}z)-\Phi(\omega^t z)\big) - \frac{\omega^{tn}}{1-\omega^{s+t}} \exp\big( \Phi(z)-\Phi(\omega^{t+s} z)\big)\right)\\
=\oint_{C_1}dz z^{n+m-1} \left( \omega^{-sn} b^{(s+t)}(\omega^{-s}z) - \omega^{tn} b^{(s+t)}(z)\right)= (\omega^{sm}- \omega^{tn}) b_{n+m}^{(s+t)}.
\end{split}
\end{align}

Proceeding in the same way in other cases we get the following commutation relations

\begin{align} \label{ab comm}
[b^{(s)}_n,b^{(t)}_m] &= (\omega^{sm}_p-\omega^{tn}_p) b^{(s+t)}_{n+m}, \qquad \text{for $s+t\neq p$,}\\
[b^{(s)}_n,b^{(p-s)}_m] &= n  \omega^{-sn}_p\delta_{n+m,0} + (\omega^{sm}_p - \omega^{-sn}_p)a_{n+m},\\
[a_n,b^{(s)}_m] &= (1-\omega^{sn}_p)b^{s}_{n+m}.
\end{align}

After defining $b^{(0)}_n \equiv a_n$, we obtain that the elements $b^{(s)}_n$ satisfy the Lepowsky-Wilson commutation relations (\ref{b comm LW}). The only difference is that in the limit of the Ding--Iohara algebra there are additional Heisenberg algebra generated by $b^{(0)}_{pn}= a_{pn}$. Therefore, the root of unity limit of the Ding--Iohara algebra has an implicit $\mathcal H \oplus \widehat{\mathfrak{sl}}(p)_1$ symmetry. This is one of the main results of our paper.

\section{AFLT basis and Macdonald polynomials limit}
In this section we construct a special basis in the representation of the $\mathcal A(1,p)$ algebra using the limit of the Ding--Iohara algebra described above. This basis is defined by the property that the vertex operators have a factorized form in it. Such a basis for the Ding--Iohara algebra was constructed in \cite{Awata:2011dc}, where it was shown that it can be described in terms of Macdonald polynomials in the sence defined below. In this paper we are interested in the root of unity limit of this construction. We will show that this basis in the Ding--Iohara algebra has a well defined limit for $q,t \rightarrow \omega_p$ and it gives a basis in the $\mathcal H \oplus \widehat{\mathfrak{sl}}(p)_1$ algebra with vertex factorization property. Macdonald polynomials with $q,t\rightarrow \omega_p$ tend to Uglov polynomials \cite{Uglov:1997ia} and the last are related to the basis in $\mathcal H \oplus \widehat{\mathfrak{sl}}(p)_1$ in the same way that Macdonald polynomials do for the Ding--Iohara algebra. This section require basic knowledge about the partitions and the Macdonald polynomials, which is gathered in Appendix A.

\subsection{Constructing a basis}
Consider the following isomorphism between the space of symmetric polynomials $\Lambda$ and the Fock space $\mathcal F_u$

\begin{align}\label{isomorph}
a_{-n} \rightarrow p_n, \quad
a_n \rightarrow n \frac{1-q^n}{1-t^n}\frac{\partial}{\partial p_n}, \text{ for } n>0,
\end{align}
where $p_n = \sum x_i^n$. This means that the level one representation of the Ding-Iohara algebra  is isomorphic to the space of symmetric functions. It is known that the Macdonald polynomials $J_\lambda(q,t)$ (see Appendix A) form a basis in the space of symmetric functions. We denote the preimages of this basis in the Fock space by the same notation $J_\lambda$. Some first examples are
\begin{align}
\begin{split}
&J_\emptyset = |u\rangle, \quad J_{(1)} = (1-t)a_{-1}|u\rangle\\
&J_{(2)} = \left(\frac 1 2 (1+q)(1-t)^2 a^2_{-1}  +\frac 1 2 (1-q)(1-t^2)a_{-2} \right)|u\rangle,\\
&J_{(1,1)} = \left(\frac 1 2 (1-t)^2(1+t)(a_{-1}^2-a_{-2})\right)|u\rangle.
\end{split}
\end{align}

The Macdonald polynomials $J_\lambda(q,t)$ in the $q,t\rightarrow e^{2\pi i/p}$ limit tends to the rank $p$ Uglov polynomials $J_\lambda ^{(\beta,p)}$ (or $\mathfrak{gl}(p)$-Jack Polynomials as it was originally called by Uglov \cite{Uglov:1997ia}). More precisely, define
\begin{equation}
J_\lambda ^{(\beta,p)}=\lim_{ \substack{q = \omega_p e^{\hbar}\\ t=\omega_p e^{\beta\hbar } \\ \hbar \rightarrow 0}}\left( \frac{J_\lambda(q,t)}{\hbar^{|\lambda^{\Diamond}|} \prod\limits_{\Box \in \lambda - \lambda^{\Diamond}}\left(1 - \omega_p^{a_\lambda(\Box)+ l_\lambda(\Box)+1}\right)} \right),
\end{equation}
where $\lambda^{\Diamond}= \{\Box \in \lambda \, |\, a_\lambda(\Box)+l_\lambda(\Box)+1 \equiv 0 \mod p\}$, $a_\lambda(\Box)$, $l_\lambda(\Box)$ are the arm and the leg of the box $\Box$.

Again Uglov polynomials form a basis in the space of symmetric functions and the isomorphism (\ref{isomorph}) with $q,t$ set to $\omega_p$ provides a relation between the space of symmetric functions and the Fock space. We denote by the same letter $J_\lambda ^{(\beta,p)}$ the preimage of the polynomial $J_\lambda ^{(\beta,p)}$ in the Fock space. Some examples of the basic vectors for $p=3$ are
\begin{align}
\begin{split}
&J_\emptyset ^{(\beta,3)}= |\kappa\rangle^\sigma, \quad J_{(1)} ^{(\beta,3)} = a_{-1}|\kappa\rangle^\sigma\\
&J_{(2)}^{(\beta,3)} = \frac 1 2 \left(  a^2_{-1}  +a_{-2} \right)|\kappa\rangle^\sigma,\\
&J_{(1,1)} ^{(\beta,3)}=\frac 1 2 \left( a_{-1}^2-a_{-2}\right)|\kappa\rangle^\sigma,\\
&J_{(3)}^{(\beta,3)} =\frac 1 2 \left(-2 \beta^{1/2} a_{-3}+3 a_{-2}a_{-1}+a_{-1}^3 \right)|\kappa\rangle^\sigma,\\
&J_{(2,1)}^{(\beta,3)} =\frac 1 2 \left(2\beta^{1/2} a_{-3}-(1+\beta)a_{-2}a_{-1} +(1-\beta)a_{-1}^3 \right)|\kappa\rangle^\sigma,\\
&J_{(1,1,1)}^{(\beta,3)} =\frac 1 2 \left(-\beta^{1/2} a_{-3}+3\beta a_{-2}a_{-1} -\beta a_{-1}^3 \right)|\kappa\rangle^\sigma,
\end{split}
\end{align}
where $|\kappa\rangle^\sigma$ is a vacuum vector of Fock space $F_\kappa^{\sigma}$ which is the limit $u\rightarrow \omega^\sigma_p e^{\kappa\hbar}$of the Fock space $F_u$ (remind that $\sigma$ distinguishes different integrable representations of $\mathcal A(1,p)$).

This basis consists of eigenvectors of an infinite set of commuting operators. This set can be obtained as a limit of the Macdonald difference operators and corresponds to the Spin Calogero--Sutherland model \cite{Uglov:1997ia,Takemura:1996qv}.

Uglov polynomials are orthogonal and their norms are given by
\begin{equation}
\langle J_\lambda^{(\beta,p)} | J_\mu ^{(\beta,p)}\rangle  = \delta_{\lambda,\mu} \omega^{|\lambda|}_p \prod_{\Box \in \lambda^{\Diamond}}\left( \beta^{1/2} l_\lambda (\Box)+\beta^{1/2} + \frac {a_\lambda(\Box)}{ \beta^{1/2} } \right) \left( \beta^{1/2} l_\lambda (\Box) + \frac{a_\lambda(\Box)+1} {\beta^{1/2}} \right).
\end{equation}
The fact that the scalar product remains indegenerate in the root of unity limit means that Uglov polynomials indeed form a basis in the integrable representation of $\mathcal H \oplus \widehat{\mathfrak{sl}}(p)_1$.

Finally we want make a notational remark. It is useful to label the basic vectors of all integrable representations of $\mathcal H \oplus \widehat{\mathfrak{sl}}(p)_1$ by colored Young diagrams. A colored Young diagram is a Young diagram colored in $p$ colors in checkerboard order. We will label it $\lambda^{\sigma}$, where $\sigma$ is the color of the corner and  its box with the coordinates $(i,j)$ has a color $i-j+\sigma \mod p$. We assign to every colored Young diagram $\lambda^\sigma$ a vector generated by Uglov polynomial $J_\lambda ^{(\beta,p)}$ in the Fock space $F_\kappa^{\sigma}$ (or in $\sigma$th integrable representation of $\mathcal H \oplus \widehat{\mathfrak{sl}}(p)_1$). This notation is quite useful since a lot of essences of the $\mathcal H \oplus \widehat{\mathfrak{sl}}(p)_1$ algebra has a simple interpretation in terms of colored Young diagrams. Examples of such interpretation are expressions for $\vec d$-grading in section \ref{vertex operators} and character formulas in section \ref{colored characters}.
\subsection{Vertex operators}\label{vertex operators}
The authors of \cite{Awata:2011dc} introduced the vertex operators $\Phi(z):\mathcal F_u \rightarrow \mathcal F_v$  for the Ding-Iohara algebra. It was shown that in the Fock realisation of the level one representation of the Ding--Iohara algebra the vertex operators have the form
\begin{equation}
\Phi(z)= \exp\left( - \sum_{n=1}^{\infty} \frac{v^n - (t/q)^nu^n}{1-q^n} \cdot \frac{a_{-n} z^n}n \right)  \exp\left( \sum_{n=1}^{\infty} \frac{v^{-n} - u^{-n}}{1-q^{-n}} \cdot \frac{a_{n} z^{-n}}n \right)
\end{equation}
In this section we will show that these operators have a well defined limit for $q,t \rightarrow \omega_p$ and they have factorized AFLT form in the Uglov basis.

Consider the following limit
\begin{equation}\label{represent limit}
q=\omega_p e^\hbar, \quad t= \omega_p e^{\beta \hbar},\quad u = \omega_p^{\sigma_1} e^{\kappa_1 \hbar},\quad v = \omega_p^{\sigma_2} e^{\kappa_2 \hbar},\quad\hbar \rightarrow 0
\end{equation}
of the vertex operator $\Phi(z)$. It approaches different limits depending on the value of $\sigma_1 - \sigma_2$. In the limit (\ref{represent limit}) we achive $p$ vertex operators acting as $\Phi^{(0)}(z): F_{\kappa_1}^{\sigma} \rightarrow F_{\kappa_2}^{\sigma}$ and $\Phi^{(k)}(z): F_{\kappa_1}^{\sigma} \rightarrow F_{\kappa_2}^{\sigma+k}$ (for any $\sigma$) corresponding to $\sigma_1=\sigma_2$ and $\sigma_1+k = \sigma_2$. By straightforward calculation we get
\begin{align}
\begin{split}
\Phi^{(0)} &= \exp\left( i(\alpha-Q)\sum_{n=1}^{\infty} \frac{a_{-pn} z^{pn}}{-pn}\right)\exp\left( i\alpha\sum_{n=1}^{\infty} \frac{a_{pn} z^{-pn}}{pn}\right)=\mathcal V_\alpha,\\
\Phi^{(k)}& = \exp\left( i(\alpha-Q)\sum_{n=1}^{\infty} \frac{a_{-pn} z^{pn}}{-pn}\right)\exp\left( i\alpha\sum_{n=1}^{\infty}  \frac{a_{pn} z^{-pn}}{pn}\right)\exp\left( -\sum_{n\not\equiv 0 \mod p} \frac{a_{n} (\omega^{k\sigma}z)^{-n}}{n}\right)=\mathcal V_\alpha \cdot \mathcal W_k,
\end{split}
\end{align}
where we use the notations $\alpha = -i \frac{\kappa_1-\kappa_2}{\sqrt \beta}$, $Q= i\left(\sqrt \beta - \frac 1 {\sqrt \beta}\right)$. $\mathcal V_\alpha$ is a rotated Heisenberg vertex operator \cite{Carlson_Okounkov:2008} and $\mathcal W_k$ is an $\widehat{\mathfrak{sl}}(p)_1$ vertex operator which permutes its integrable representations.

To find matrix element of these vertex operators, we present the following formula for the matrix elements of the vertex operators of the level one representation of the Ding--Iohara algebra in the basis $J_\lambda$ \cite{Awata:2011dc}
\begin{equation}
\langle J_\lambda|\Phi(z)|J_\mu \rangle = N_{\lambda,\mu} \left(\frac{qv}{tu}\right)  \left(\frac{tu}{q}\right)^{|\lambda|}  \left(-\frac{v}{q}\right)^{-|\mu|}t^{n(\lambda)}q^{n(\mu')} z^{|\lambda|-|\mu|},
\end{equation}
where $n(\lambda) = \sum_{i\ge 1} (i-1)\lambda_i$, $\mu'$ is partition transposed to $\mu$ and
\begin{equation}
N_{\mu,\nu}=\prod_{\Box \in \lambda} (1-uq^{-a_\mu(\Box)-1}t^{-l_\lambda(\Box)})\cdot\prod_{\Box \in \mu} (1-uq^{a_\lambda(\Box)}t^{-l_\mu(\Box)+1}),
\end{equation}
where $a_\lambda(\Box)$ and $l_\lambda(\Box)$ are arm and leg of the box $\Box$.

For the case of $\sigma_1=\sigma_2=\sigma$ in the root of unity limit the matrix elements become
\begin{align}
\begin{split}
\langle J_{\lambda^{\sigma}}^{(\beta,p)}| \Phi^{(0)}(z)| J_{\mu^\sigma}^{(\beta,p)}\rangle &=(-1)^{|\mu|} \omega_p^{n(\lambda)+n(\mu)+\sigma |\lambda|-(\sigma-1)|\mu|}z^{|\lambda|- |\mu|}N^{(\beta,p)}_{\lambda^{\sigma},\mu^{\sigma}}(\alpha), \quad  \text{if }  \vec d(\lambda^{\sigma}) =  \vec d (\mu^{\sigma}),
\end{split}
\end{align}
and the matrix element is zero otherwise. We used notations  $\vec d(\lambda^{\sigma})$ for a vector with components $(E^{(i,i)}_0-E^{(i+1,i+1)}_0) (\lambda^{\sigma})$ where $i=1,\dots, p-1$ and $(E^{(i,i)}_0-E^{(i+1,i+1)}_0) (\lambda^{\sigma})$ is  $E^{(i,i)}_0-E^{(i+1,i+1)}_0$~-grading of the vector $|\lambda^{\sigma}\rangle$. We also used
\begin{equation}
N^{(\beta,p)}_{\lambda^{\sigma_1},\mu^{\sigma_2}}(\alpha) = \prod_{\Box\in S(\lambda^{\sigma_1},\mu^{\sigma_2})}\left(1 - i\alpha\sqrt \beta + a_\lambda(\Box)+\beta l_\lambda(\Box)\right)\prod_{\Box\in S(\mu^{\sigma_2},\lambda^{\sigma_1})}\left(1-i \alpha\sqrt \beta + a_\mu(\Box)+\beta l_\mu(\Box)\right),
\end{equation}
where $\Box \in S(\lambda^{\sigma_1},\mu^{\sigma_2})$ if $\Box \in \lambda$ and $l_\lambda(\Box)+a_\mu(\Box)+\sigma_2-\sigma_1 \equiv 0 \mod p$.

For the case of $\sigma_1 = \sigma_2+k$ the root of unity limit of the matrix elements become

\begin{align}
\langle J_{\lambda^{\sigma_1}}^{(\beta,p)}| \Phi^{(k)}(z)| J_{\mu^{\sigma_2}}^{(\beta,p)}\rangle &=(-1)^{|\mu|} \omega_p^{n(\lambda)+n(\mu)+\sigma_1 |\lambda|-(\sigma_2-1)|\mu|}z^{|\lambda|- |\mu|}N^{(\beta,p)}_{\lambda^{\sigma_1},\mu^{\sigma_2}}(\alpha),\\ \label{condition}\quad  \text{if }  \vec d(\lambda^{\sigma_1}) &=  \vec d (\mu^{\sigma_2})+\vec d(\rho)\quad \text{for some } \rho \in \mathsf{F}_k (\mathfrak{sl}(p)),
\end{align}
where $\rho$ is one of the weights of the $k$-th fundamental representation  $\mathsf{F}_k (\mathfrak{sl}(p))$ of $\mathfrak{sl}(p)$ and $\vec d(\rho)$ is a vector of its  $E^{(i,i)}_0-E^{(i+1,i+1)}_0$~-gradings. When condition (\ref{condition}) is not satisfied the matrix element is zero.

We also note that $\vec d(\lambda^{\sigma})$ can be counted directly from a partition $\lambda^{\sigma}$ by the formula
\begin{equation}
d_i(\lambda^{\sigma}) = (E^{(i,i)}_0-E^{(i+1,i+1)}_0) (\lambda^{\sigma}) = 2N_{p-i-1}(\lambda^{\sigma}) - N_{p-i}(\lambda^{\sigma})-N_{p-i-2}(\lambda^{\sigma}) - \delta_{\sigma+i+1,p},
\end{equation}
where $N_i(\lambda^{\sigma})$ is a number of the boxes of $\lambda^{\sigma}$ colored in $i$-th color.
\section{Character formulas}
The fact that $\mathcal H \oplus\hat{\mathfrak{sl}}(p)_1 $ can be realised in terms of its  Heisenberg subalgebra provide us with a couple of non-trivial relations for the characters of the integrable representations of the $\mathcal H \oplus \widehat{\mathfrak{sl}}(p)_1$. These formulas are specification of Macdonald identities for characters of the affine algebras, but we think that our interpretation of them are of interest.

\subsection{Case $p=2$}
The homogeneous grading in  the   $\hat{\mathfrak{sl}}(2)_1$ algebra is defined by two operators $L_0$ and $h_0/2$, where $L_0$ acts as
$$
[L_0, e_{-n}] = ne_{-n}, [L_0, h_{-n}] = nh_{-n}, [L_0,f_{-n}] = nf_{-n}.
$$

Here we use the same notations as in the section 2 for $\hat{\mathfrak{sl}}(2)_1$. The formula for the homogeneous character of the representation is
\begin{equation}
\chi_H(q,t) = \text{Tr}_{\pi_{\sigma,1}} q^{L_0}t^{h_0/2 }.
\end{equation}
The Lepowsky-Wilson construction breaks both $L_0$ and $h_0/2$ gradings. For example, one can see that operator $a_{2n+1}=e_n+f_{n+1}$ has neither definite $L_0$ nor $h_0/2$ grade. But it preserves principal grading defined by operator $\text{Pr} = L_0 - h_0/4$. The principal character of the representation is defined as
\begin{equation}\label{principal character}
 \chi_P(q,t) = \text{Tr}_{\pi_{\sigma,1}} q^{\text{Pr}}t^{h_0/2 }.
\end{equation}
Since the  level one representation of $\hat{\mathfrak{sl}}(2)_1$ can be realised as a Fock module $\mathbb C[a_{-1},a_{-3},\dots]$ we can write\footnote{we drop even modes $a_{2n}$ since they correspond to $\mathcal H$ in $\mathcal H\oplus\hat{\mathfrak{sl}}(2)_1$ and completely decouple}
\begin{align}\label{char 2}
\chi_P(q) = \prod_{n=1}^{\infty} \frac{1}{1-{q^{n-1/2}}},
\end{align}
where the right-hand is the principal character of a Fock module $\mathbb C[a_{-1},a_{-3},\dots]$ and $\chi_P(q)$ means (\ref{principal character}) with $t=1$. On the other hand the principal character of $\hat{\mathfrak{sl}}(2)_1$ is known to be

$$
\chi_{P}(q,t) = \chi_B(q)\sum_{n\in\mathbb Z} q^{n^2+n/2} t^n,
$$
where $\chi_B(q)=1/\prod_{n=1}^{\infty} (1-q^n)$ is a bosonic character. Therefore we have a relation
\begin{equation}
\chi_B(q)\sum_{n\in\mathbb Z} q^{n^2+n/2}=\prod_{n=1}^{\infty} \frac{1}{1-{q^{n-1/2}}}.
\end{equation}
We derived it using the fact that $\hat{\mathfrak{sl}}(2)_1$ can be realised in terms of its Heisenberg subalgebra $a_{2n+1}$. However, it could be proven directly by use of triple Jacobi identity

\begin{align}
\sum_{n\in \mathbb Z} x^{n^2}y^{n} = \prod_{m=1}^\infty(1-x^{2m})(1+x^{2m-1}y)(1+x^{2m-1}y^{-1}) .
\end{align}
Setting $x=q,y={q^{1/2}}$ we have
\begin{align}
\chi_B(q)\sum_{n\in \mathbb Z} q^{n^2+n/2} = \prod_{m=1}^{\infty}\frac{(1-q^{2m})(1+q^{2m-1/2})(1+q^{2m-3/2})}{(1-q^{m})}= \prod_{m=1}^{\infty}\frac{1}{1-q^{m-1/2}}.
\end{align}

\subsection{Case $p=3$}
Previous idea can be extended to the cases of the general $p$. We will present some results for the $p=3$ case
In this case the principal character is
\begin{align}
\chi_P (q)= \chi_B(q)^2\sum_{k_1,k_2\in \mathbb Z} q^{k_1^2+k_2^2-k_1k_2+\frac{k_1+k_2}3 - k_r}.
\end{align}
It is independent of the representation number $r$. And since $\hat{\mathfrak{sl}}(3)_1$ can be constructed in the terms of a Fock module $\mathbb C[a_{-1},a_{-2},a_{-4},a_{-5},a_{-7}, \dots]$ we have
\begin{align}
\chi_B(q)^2\sum_{k_1,k_2\in \mathbb Z} q^{k_1^2+k_2^2-k_1k_2+\frac{k_1+k_2}3 - k_r} = \prod_{n=1}^{\infty} \frac{1}{(1-q^{n-1/3})(1-q^{n-2/3})}
\end{align}
It could be proved directly by use of the Macdonald identity for $\hat{\mathfrak{sl}}(3)_1$ algebra.

\subsection{Characters and colored Young diagrams} \label{colored characters}
As we had shown in section 4.1, the basis of integrable representation consists of vectors labeled by colored Young diagrams (with fixed color in the corner of the diagram). Therefore, there should be relation between the generating function of the colored Young diagrams and the character of the integrable representation of $\mathcal H\oplus\hat{\mathfrak{sl}}(p)_1$. We define the generating function of the colored Young diagrams as \cite{Alfimov:2013cqa}
\begin{align}
\begin{split}
&\chi_Y^{(r)}(q,k_1,\dots,k_{p-1})\\&=\sum_{\lambda} \#\big\{\text{Young diagrams $\lambda$ with $r$-coloured corner} \mathbin | N_i(\lambda^r)- N_0(\lambda^{r})=k_i, \text{ for all $i$}\big\} \,q^{|\lambda|/p},
\end{split}
\end{align}
where $\#$ means "number of", and $N_i(\lambda^{r})$ means number of boxes of $\lambda^r$ colored in $i$-color. In other words, the generating function of Young diagrams counts the number of Young diagram with fixed color in the corner, fixed number of boxes and fixed differences between the number of boxes with fixed color. One can show that it is equal to the prinicpal character of the integrable representation of $\mathcal H\oplus\hat{\mathfrak{sl}}(p)_1$
\begin{align}
\begin{split}
\sum_{k_i\in \mathbb Z}\chi_Y^{(r)}(q,k_1,\dots,k_{p-1}) t_1^{k_1}\dots t_{p-1}^{k_{p-1}}= \chi_P^{\mathcal H\oplus\hat{\mathfrak{sl}}(p)_1}(q,t_1,t_2,\dots,t_{p-1}) \\= \sum_{k_i\in \mathbb Z} \chi_B(q)^p q^{\sum_{i=1}^p \left( k_i^2-k_ik_{i+1}+\frac{k_i}{p} \right)-k_r}t_1^{k_1}\dots t_{p-1}^{k_{p-1}}.
\end{split}
\end{align}
It is interesting that Lepowsky-Wilson construction breaks $\vec d(\lambda^\sigma)$-grading, but Uglov basis still has a definite $\vec d$-grade.
\section{Conclusion}
In the present paper we explicitly constructed the $\mathcal H \oplus \widehat{\mathfrak{sl}}(p)_1$ algebra from the limit of the Ding--Iohara algebra and we showed that there is a special basis in which vertex operators have a simple factorized form. These calculation support the conjecture that AGT correspondence between the $U(r)$ instanton counting on $\mathbb R^4/\mathbb Z_p$ and the two-dimensional field theory with symmetry $\mathcal A(r,p)$ can be considered as the root of unity limit of the  five-dimensional AGT relation. It is interesting that such a set of different relations can be considered as a limit of a single one.

{\bf{Acknowledgements.}} Author is grateful to A. Belavin, Ya.Pugai,  M. Lahskevich and  M. Bershtein for useful comments and discussions. The research was performed under a grant funded by Russian Science Foundation (project
No. 14-12-01383).

\appendix

\section{Appendix A: Partitions and Macdonald polynomials}
A partitions $\lambda$ is a set of nonnegative integers $(\lambda_1,\lambda_2,\dots)$ such that there are only finite number of nonzero entries and $\lambda_1\ge \lambda_2 \ge \dots$. Length $l(\lambda)$ is number of nonzero entries, $|\lambda| = \sum_i \lambda_i$. The conjugate partition $\lambda'$ correspond to the transpose of the diagram $\lambda$. The dominance ordering $\lambda>\mu$ means $|\lambda|= |\mu|$ and $\sum_{i=1}^k \lambda_i > \sum_{i=1}^k \mu_i$ for any $k$. We use notation $\square = (i,j)$ for labeling a box located at the coordinate $(i,j)$. Arms and legs of the box are defined as
\begin{equation}
a(\square) = \lambda_i - j, \quad l(\square) = \lambda'_j-i.
\end{equation}

 We use notations $p_\lambda = \prod_{i=1}^{l(\lambda)}p_{\lambda_i} $, where   $p_k= \sum x^k_i$ for the power-sum symmetric functions and $m_\lambda = \text{Sym\,} \prod_{i=1}^{l(\lambda)}\left( x_i^{\lambda_i}\right)$ for the monomial symmetric function.

The Macdonald polynomials $P_\lambda(q,t)$ are uniquely defined by two properties. They are orthogonal under scalar product
\begin{equation}
\langle p_\lambda, p_\mu\rangle = \delta_{\lambda,\mu} z_\lambda \prod_{i=1}^{l(\lambda)} \frac{1-q^\lambda_i}{1-t^{\lambda_i}}, \quad z_\lambda = \prod_{i\ge 1}i^{m_i} m_i!,
\end{equation}
and the transitions function between the basis $m_\lambda$ and the basis of $P_\lambda(q,t)$ is upper unitriangular
\begin{equation}
P_\lambda(q,t) = m_\lambda + \sum_{\mu<\lambda} u_{\lambda,\mu}(q,t)m_\mu,
\end{equation}
where $\mu<\lambda$ is the dominance order. Polinomials $P_\lambda(q,t)$ have the normalization
\begin{equation}
\langle P_\lambda(q,t), P_\lambda(q,t) \rangle = \prod_{\square \in \lambda} \frac{1-q^{a(\square)+1}t^{l(\square)}}{1-q^{a(\square)}t^{l(\square)+1}},
\end{equation}
where $a(\square)$ and $l(\square)$ are arms and legs of box $\square$ and product is over all boxes of $\lambda$.

The integral form $J_\lambda(q,t)$ of Macdonald polynomials is defined as
\begin{equation}
J_\lambda(q,t) = c_\lambda(q,t) P_\lambda(q,t), \quad \text{where } c_\lambda(q,t) = \prod_{\square \in \lambda}(1-q^{a(\square)}t^{l(\square)+1}).
\end{equation}
They have norms
\begin{equation}
\langle J_\lambda(q,t), J_\lambda(q,t) \rangle = \prod_{\square \in \lambda}(1-q^{a(\square)}t^{l(\square)+1})(1-q^{a(\square)+1}t^{l(\square)}).
\end{equation}

\end{document}